\documentclass[%
 reprint,
 amsmath,amssymb,
 aps,
prb,
superscriptaddress,
]{revtex4-1}

\usepackage{graphicx}
\usepackage{dcolumn}
\usepackage{bm}
\usepackage{hhline}
\usepackage[utf8]{inputenc}
\usepackage[english]{babel}
\usepackage{color}

\begin{document}

\preprint{APS/123-QED}


\title{FeTaSb and FeMnTiSb as promising thermoelectric materials: An \textit{ab initio} approach}

\author{Mohd~Zeeshan}
\affiliation{Indian Institute of Technology Roorkee, Department of Chemistry, Roorkee 247667, Uttarakhand, India}
\author{Tashi Nautiyal}
\affiliation{Indian Institute of Technology Roorkee, Department of Physics, Roorkee 247667, Uttarakhand, India}
\author{Jeroen van den Brink}
\affiliation{Institute for Theoretical Solid State Physics, IFW Dresden, Helmholtzstrasse 20, 01069 Dresden, Germany}
\author{Hem C. Kandpal}
\affiliation{Indian Institute of Technology Roorkee, Department of Chemistry, Roorkee 247667, Uttarakhand, India}

\date{\today}

\begin{abstract}
Thermoelectricity in principle provides a pathway to put waste heat to good use. Motivated by this we investigate thermal and electrical transport properties of two new Fe-based Heusler alloys, FeTaSb and FeMnTiSb, by a first principles approach and semiclassical Boltzmann transport theory within the constant relaxation-time approximation. We find a high power factor of \textit{p}-doped FeTaSb, competitive with best performing Heusler alloy FeNbSb at 1100 K. The obtained power factor of \textit{n}-doped FeMnTiSb at room temperature is higher than that of both FeNbSb and FeTaSb. Remarkably, FeMnTiSb can be used for both \textit{n}-type and \textit{p}-type legs in a thermoelectric module. The Seebeck coefficients of the two proposed systems are in line with those of earlier reported Heusler alloys. We also provide conservative estimates of the figure of merit for the two systems. Overall, our findings suggest a high temperature thermoelectric potential of FeTaSb while the low cost FeMnTiSb is a viable room temperature thermoelectric candidate material. 

\begin{description}
\item[Usage]
Secondary publications and information retrieval purposes.
\item[PACS numbers]
May be entered using the \verb+\pacs{#1}+ command.
\item[Structure]
You may use the \texttt{description} environment to structure your abstract;
use the optional argument of the \verb+\item+ command to give the category of each item. 
\end{description}
\end{abstract}

\pacs{Valid PACS appear here}
\maketitle


\section{Introduction}
Thermoelectric (TE) materials can in principle utilize waste heat to provide an alternate source of power generation \cite{Samanta17, Lee17, Zhao17}. However, due to low efficiency, TE materials still only find niche applications \cite{He17, Boona17}. The application potential is limited because the efficiency of a TE material is governed by a generally conflicting requirement on the physical properties of a material: the electrical conductivity must be high while at the same time thermal conductivity should be low. The efficiency is measured by the dimensionless figure of merit \textit{ZT} = \textit{S$^2\sigma$T/$\kappa$}, where \textit{S} is Seebeck coefficient, $\sigma$ is electrical conductivity, \textit{T} is absolute temperature, \textit{$\kappa$} = \textit{$\kappa_e$} + \textit{$\kappa_l$} and \textit{$\kappa_e$} and \textit{$\kappa_l$} are electronic and lattice components of thermal conductivity \cite{Tan17, Ge17, Huang17, Sevincli13}. 
Another issue is the cost of materials since the best performing TE systems have expensive constituents such as Hf, Te, and Ge \cite{Rai17, Joshi14, He14, LeBanc14}. Therefore in the search for materials with a higher \textit{ZT}, factors like nontoxicity and earth crust abundance also play an important role \cite{Yabuuchi17, Schierning15}.

According to Slack, a good thermoelectric material should have {\it phonon-glass electron-crystal} behavior \cite{Slack92, Slack95}. The concept suggests that for a high \textit{ZT}, the material should possess electronic properties similar to a crystal and thermal properties as that of an amorphous glass. Obviously, it is not easy to design such a material, nevertheless, the criterion is most closely followed by narrow band gap semiconductors, comprising of heavy elements. Heusler alloys are found to be in line with these requirements. They have good thermal stability, mechanical strength, low cost, nontoxicity, and semiconducting behavior \cite{Misra14, Xie12, Young00, Culp08, Casper12}. Good thermal stability and mechanical strength make them in principle suitable for high temperature thermoelectric applications.  

As per Slater-Pauling rule, most of the 18 valence electron count (VEC) half-Heusler (hH) alloys, \textit{XYZ}, and 24 VEC full Heusler alloys, \textit{X$_2$YZ}, or quaternary Heusler alloys, \textit{XX$^\prime$YZ}, are found to be narrow band gap semiconductors \cite{Kubler84, Galanakis02, Felser07, Ozdogan13}. Also, most Heusler alloys are less expensive in comparison to most of the state-of-art TE materials. For instance, TiNiSn, TiCoSb, FeVSb, FeNbSb, Fe$_2$VAl, and Fe$_2$TiSi contain predominantly abundant earth crust elements \cite{Jong17}. Despite these advantages, the high thermal conductivity of many Heusler alloys limits their applicability \cite{He16, Carrete14, Snyder08}.

Recently, many Heusler alloys have been found to exhibit good TE properties comparable to conventional Bi$_2$Te$_3$ and PbTe based TE materials \cite{Fang17, Yan12, Yu09, Sakurada05}. \textit{M}NiSn and \textit{M}CoSb (where M = Ti, Zr, Hf) are the two pioneer and most studied hH classes \cite{Berry17, Chen13}. A high \textit{ZT} of 1.05 and 0.8 near 900-1000 K has been reported for the \textit{n}-type Hf$_{0.6}$Zr$_{0.4}$NiSn$_{0.995}$Sb$_{0.005}$ hH alloy and \textit{p}-type Hf$_{0.3}$Zr$_{0.7}$CoSn$_{0.3}$Sb$_{0.7}$/nano-ZrO$_2$ composites, respectively \cite{Poon11}. But the drawback of these materials is the expensive Hf as the major constituent. The quest for higher \textit{ZT} and low cost materials shifted the focus towards a new promising hH class \textit{M}FeSb (\textit{M} = V, Nb, and Ta). The 5\textit{d} element Ta, though expensive, can be used for doping a small amount in FeVSb or FeNbSb to improve the TE properties.

Within a short span of time, a lot of work has been reported in this family \cite{Feng16, Coban15, Zhu15, Zhu13, Zou13}. A high power factor (PF), \textit{$S^2\sigma$}, of 48 $\mu$W cm$^{-1}$ K$^{-2}$ is reported for FeVSb system. Despite a high PF, the maximum reported \textit{ZT} is 0.25 owing to a high thermal conductivity of FeVSb \cite{Fu13}. Doping attempts record the highest \textit{ZT} of 0.8\cite{Fu14}. More interestingly, FeNbSb based hH alloys have been reported to exhibit \textit{ZT} $>$ 1 \cite{Fu15, Fu16}. A record peak \textit{ZT} of 1.6 is achieved for Nb$_{0.6}$Ta$_{0.4}$Ti$_{0.2}$FeSb at 1200 K \cite{Yu17}. The least explored member FeTaSb of the family has a patent filed in 2015 on a theoretical approach \cite{Patent}. Recently, it has been predicted as a \textit{p}-type hH candidate \cite{Bhattacharya16}.

The figure of merit of Heusler alloys has seen a progressive increase in past two decades but has yet to touch the threshold value for practical applications \cite{Bell08, Biswas12}. Over the years, various strategies have been adopted to improve the efficiency of TE materials. Two broad approaches are i) enhancing the PF and ii) minimizing the thermal conductivity. The PF can be improved by band gap engineering and controlling the charge carrier concentration through doping. Iso-electronic alloying, doping or nanostructuring could be helpful in reducing the thermal conductivity \cite{Yu17, Dehkordi15, Chen13, Heremans08, Snyder08}. 

Further, there is a surge in efforts for developing new materials with lower thermal conductivity \cite{Liu16}. The interplay of theory and experiment has been fruitful in the past for designing new materials \cite{Yang16, Gautier15, Zakutayev13}. The theoretical prediction of stable systems allows the experimentalists to narrow down the window for targeting new materials. Likewise, the theoretical prediction of optimal doping levels for attaining better TE efficiency has been very useful \cite{Yang08, Yabuuchi13, Zeeshan17}. The \textit{ab initio} approach, semiclassical Boltzmann theory, and rigid band approximation have recently been used as effective tools for predicting the TE properties \cite{Shi17, Parker10, Yang08, Madsen06JACS, Zeeshan17, Zeeshan2017}. 

In the present work, we use \textit{ab initio} electronic structure calculations, semiclassical Boltzmann transport theory, and a rigid band approximation to systematically investigate the thermodynamic stability, dynamical stability, and thermal and electrical transport properties of FeTaSb and FeMnTiSb in cubic F$\bar{4}$3m symmetry. The custom choice of materials is based on earlier discussion and fulfils the foremost criterion of semiconducting behavior. Also, the FeMnTiSb constitutes low cost elements. On descending in \textit{M}FeSb family from V to Ta, Ta being heaviest is expected to scatter the phonons effectively on account of mass difference \cite{Chakraborty17}. As FeNbSb is reported to have a high \textit{ZT}, we have high hopes from FeTaSb. This is because of the similar sizes of Nb and Ta that their electrical properties are expected to be unaltered. Further, we consider the doping effect on TE properties of FeTaSb. 
 
Conventionally, doping amounts to replacing a certain amount of an atom by a desirable dopant e.g., FeVSb to FeV$_{0.6}$Nb$_{0.4}$Sb \cite{Fu12} or doping a certain amount in the void spaces. As the hH structure has the advantage of vacant lattice sites, instead of doping a small fixed amount, we choose to introduce a fourth atom into all available vacant lattice sites. This leads to the possibility of making FeTaSb a full-Heusler by doping Fe or Ta e.g., Fe$_2$TaSb. However, this compound does not obey the 24 VEC Slater-Pauling criterion of a semiconductor. Introducing a different fourth atom may lead to effective phonon scattering on account of mass fluctuations \cite{Fu12} and as well may satisfy the Slater-Pauling rule. After analyzing different possible combinations in order to obey the Slater-Pauling rule with an introduction of the fourth atom in the crystal structure, we arrive at FeMnTiSb. 

The hypothetical quaternary Heusler alloy FeMnTiSb fulfils the 24 VEC semiconductor criterion, comprises of low cost elements, and is expected to have lower thermal conductivity on account of more scattering points and mass fluctuations. Moreover, the choice of Mn as the fourth atom is also suitable on account of bulk low thermal conductivity. Scanning of the literature reveals that there is no prior experimental evidence for existence or synthesis of both FeTaSb and FeMnTiSb. Also, quaternary Heusler alloys have not been much explored in the context of thermoelectric materials.

The present paper is organized as follows. Section II includes a brief description of the computational tools used for the study and the crystal structure of FeTaSb and FeMnTiSb in cubic F$\bar{4}$3m symmetry. In Sec. III, we discuss structural optimization, ground state properties, static and dynamic stability (phonons),  electronic structure (band structure and DOS), thermal and electrical transport properties, and their behavior with optimal doping at different temperatures. In Sec. IV, we summarize the important observations of the work. We use FeNbSb, an experimentally well-explored system, as a reference to check the compatibility of our results. 

\begin{figure}[h]
\centering
\includegraphics[scale=0.35]{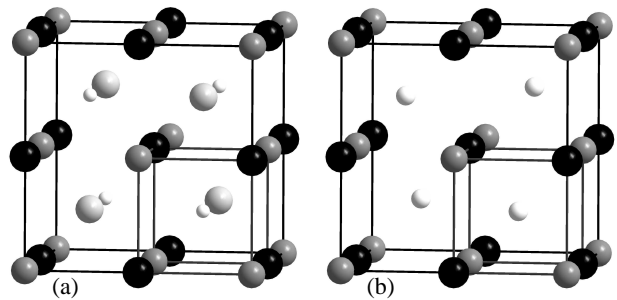}
\caption{The crystal structure of (a) quaternary and (b) half-Heusler alloy in cubic F$\bar{4}$3m symmetry. White, light gray, dark gray, and black spheres represent \textit{X}, \textit{X$^\prime$}, \textit{Y}, and \textit{Z}, respectively. The spheres are shown of different sizes for clarity.}
\label{struct}
\end{figure}

\section{Computational Details and Crystal Structure}
We use a combination of two different first-principles density functional theory (DFT) codes: the full-potential linear augmented plane wave method (FLAPW)\cite{Singh06} implemented in WIEN2K\cite{Blaha01} and the plane-wave pseudopotential approach implemented in QUANTUM ESPRESSO package\cite{Giannozzi09}. The former has been used to obtain equilibrium lattice constants, electronic structure, and transport properties, and the latter to confirm the structure stability by determining the phonon spectrum.

The FLAPW calculations are performed using a modified Perdew-Burke-Ernzerhof (PBEsol correlation)\cite{Perdew08} implementation of the generalized gradient approximation (GGA). For all the calculations, the scalar relativistic approximation is used. The muffin-tin radii (RMTs) are taken in the range 2.3--2.6 Bohr radii for all the atoms. RMT $\times$ kmax = 7 is used as the plane wave cutoff. The self-consistent calculations were employed using 125000 \textit{k}-points in the full Brillouin zone. The energy and charge convergence criterion was set to 10$^{-6}$ Ry and 10$^{-5}$ e, respectively. 

The electrical transport properties have been calculated using the Boltzmann theory\cite{Allen} and relaxation time approximation as implemented in the BOLTZTRAP code\cite{Madsen06}. The electrical conductivity and power factor are calculated with respect to time relaxation, \textit{$\tau$}; the Seebeck coefficient is independent of \textit{$\tau$}. The relaxation time was calculated by fitting the theoretical data with available experimental data. 


In the plane-wave pseudopotential approach, we use scalar-relativistic, norm-conserving pseudopotentials for a plane-wave cutoff energy of 100~Ry. The exchange-correlation energy functional was evaluated within the GGA, using the Perdew-Burke-Ernzerhof parametrization\cite{Perdew96}, and the Brillouin zone is sampled with a 20$\times$20$\times$20 mesh of Monkhorst-Pack \textit{k}-points. The calculations are performed on a 2$\times$2$\times$2 \textit{q}-mesh in the phonon Brillouin zone.  

We obtain the lattice thermal conductivity by solving linearized Boltzmann transport equation (BTE) within the single-mode relaxation time approximation (SMA)\cite{Ziman1960} using thermal2\cite{thermal2} code as implemented in the QUANTUM ESPRESSO package. Further, we have chosen Troullier-Martins norm-conserving pseudopotentials from the QUANTUM ESPRESSO webpage\cite{QEweb}. 

The crystal structure of quaternary Heusler alloy \textit{XX$^\prime$YZ} can be visualized as a CsCl superstructure. If the lattice parameter of CsCl unit cell is doubled along all three axes, a cell with 8 cubes having one atom at the center of each cube is obtained. The atoms \textit{Y} and \textit{Z} occupy corners whereas \textit{X} and \textit{X$^\prime$} atoms occupy the center of alternate cubes \cite{Claudia_Book, Graf11}. If \textit{X$^\prime$} atoms are missing from the lattice, the resulting structure leads to the hH alloy, \textit{XYZ}. Both \textit{XYZ} and \textit{XX$^\prime$YZ} crystallize in F$\bar{4}$3m symmetry. 

The Wyckoff positions for \textit{X}, \textit{Y}, and \textit{Z} in hH alloy are (1/4, 1/4, 1/4), (0, 0, 0), and (1/2, 1/2, 1/2), respectively, with vacant positions at (3/4, 3/4, 3/4) whereas the Wyckoff positions for \textit{X}, \textit{X$^\prime$}, \textit{Y}, and \textit{Z} in quaternary Heusler alloy are (3/4, 3/4, 3/4), (1/4, 1/4, 1/4), (0, 0, 0), and (1/2, 1/2, 1/2), respectively \cite{Xiong14}. \textit{X}, \textit{X$^\prime$}, and \textit{Y} are transition elements whereas \textit{Z} comes from the main group elements and atoms are arranged in a face-centered pattern (Fig.~\ref{struct}).

\section{Results and Discussion}

\subsection{Structural Optimization and Stability}

\begin{figure}
\centering
 \includegraphics[scale=0.4]{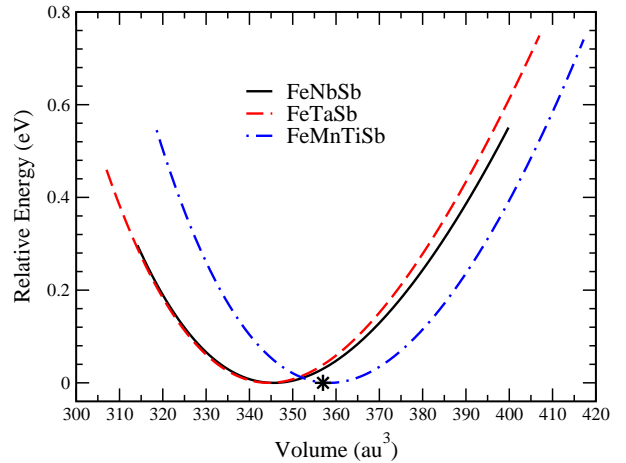}
 \caption{Calculated relative total energy as a function of volume for FeNbSb, FeTaSb, and FeMnTiSb in cubic F$\bar{4}$3m symmetry. The experimentally reported volume of FeNbSb is marked by asterisk\cite{Melnyk00}. For better illustration, the energy minimum of all plots is taken as zero on energy axis.}
\label{optimize}
\end{figure}

In order to determine the ground state properties, we optimized the crystal structures of FeNbSb, FeTaSb, and FeMnTiSb using GGA-PBEsol implemented in WIEN2K. The ground state of all the three systems was found to be nonmagnetic. We minimized the total energy as a function of unit cell volume, fitted with Birch-Murnaghan equation \cite{Birch47}, in F$\bar{4}$3m symmetry (Fig.~\ref{optimize}). There is a slight deviation in calculated and experimental volume of FeNbSb \cite{Melnyk00}. However, the energy well of FeTaSb is quite similar to that of FeNbSb, indicating the possibility of the existence of FeTaSb too. Furthermore, the energy profile of FeMnTiSb is also similar though somewhat shifted towards the higher volume due to the presence of the fourth atom in the unit cell. The considered systems thus can be synthesized under suitable conditions.

The lattice parameter and the band gap of all the systems are listed in Table~\ref{lattice}. The calculated and experimental lattice parameters of FeNbSb are in agreement with a discrepancy of only 0.92\%\/. The lattice parameter of FeTaSb is slightly lower than that of FeNbSb \cite{Housecroft_Book} which can be attributed to the effect of lanthanide contraction on post lanthanides. Though FeMnTiSb has not been reported yet, FeMnTiAs was studied in a previous \textit{ab initio} electronic structure calculation with lattice parameter as 5.79~\AA\/ and its ground state was found to be nonmagnetic \cite{Ozdogan13}. For the purpose of comparison, the calculated lattice parameter of FeMnTiSb is larger than that of FeMnTiAs, as expected, due to the size difference between Sb and As. 

As expected, all the systems are nonmagnetic semiconductors in accordance with the Slater-Pauling rule. To the best of our knowledge, there is no experimental report on the band gap of FeNbSb. Our calculated band gap of FeNbSb is in good agreement with previously calculated values \cite{Hong16, Fang16, Joshi14}. Going from FeNbSb to FeTaSb, the increase in band gap can be explained in terms of electronegativity difference of Nb and Ta \cite{Graf11}. The large band gap of FeTaSb, in comparison to FeNbSb, could be helpful in suppressing the onset of bipolar conduction at high temperature, thereby improving the TE properties at elevated temperatures\cite{Yu17}. FeMnTiSb has the smallest band gap among all the systems since the close proximity of atoms allows the larger overlap and more delocalization of the orbitals\cite{Tobola00}. 

\begin{table}[]
\centering
\setlength{\arrayrulewidth}{1pt}
\begin{tabular}{c|c|c}
\hline 
\hline 
System   & a (\AA)      		 & E$_g$ (eV)  			\\ \hline
FeNbSb   & 5.8950 (5.95)\cite{Melnyk00}  & 0.53 (0.53)\cite{Joshi14}    \\
FeTaSb   & 5.8879       		 & 0.86        			\\
FeMnTiSb & 5.9769       	 	 & 0.31        			\\ \hline
\hline 
\end{tabular}
\caption{The calculated lattice parameter $a$ and band gap E$_g$ values of all the systems in cubic F$\bar{4}$3m symmetry. The reported values are given in parentheses.}
\label{lattice}
\end{table}

The optimized structures of FeNbSb, FeTaSb, and FeMnTiSb are further studied for dynamic stability with phonon calculations. We performed a two-step phonon calculation. First, the crystal structure was optimized by using QUANTUM ESPRESSO, based on DFT and plane-wave pseudopotential method. The optimized results were in good agreement with our WIEN2K findings. Then, we calculated the phonon dispersion by using the density functional perturbation theory (DFPT) implemented in QUANTUM ESPRESSO. The calculations were performed on a 2$\times$2$\times$2 mesh in the phonon Brillouin zone, and force constants in real space derived from this input are used to interpolate between q-points and to obtain the continuous branches of the phonon band structure. 
 
\begin{figure}
\centering
 \includegraphics[scale=0.4]{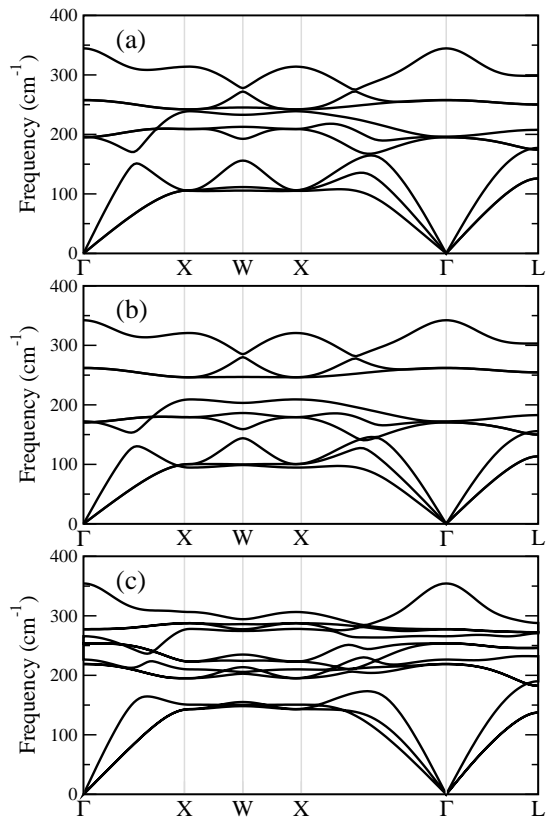}
 \caption{The phonon dispersion curves of (a) FeNbSb, (b) FeTaSb, and (c) FeMnTiSb, in cubic F$\bar{4}$3m symmetry.}
 \label{phonon}
\end{figure}

Phonons are often regarded as normal modes or quantum of vibrations in a crystal. For a system to be dynamically stable, the frequency of each phonon should be real and not imaginary \cite{Elliott06, Togo15}. Our results show no imaginary frequencies for all the systems throughout the Brillouin zone, thus, ensuring the stability of FeNbSb, FeTaSb, and FeMnTiSb. As can be seen from Fig.~\ref{phonon}, there are three acoustical and six optical branches for FeNbSb and FeTaSb whereas three acoustical and nine optical branches for FeMnTiSb. The major contribution to lattice thermal conductivity comes from the acoustical phonon branches since they have large group velocities \cite{Pichanusakorn10}. 

\subsection{Electronic Structure}
\begin{figure}
\centering
 \includegraphics[scale=0.4]{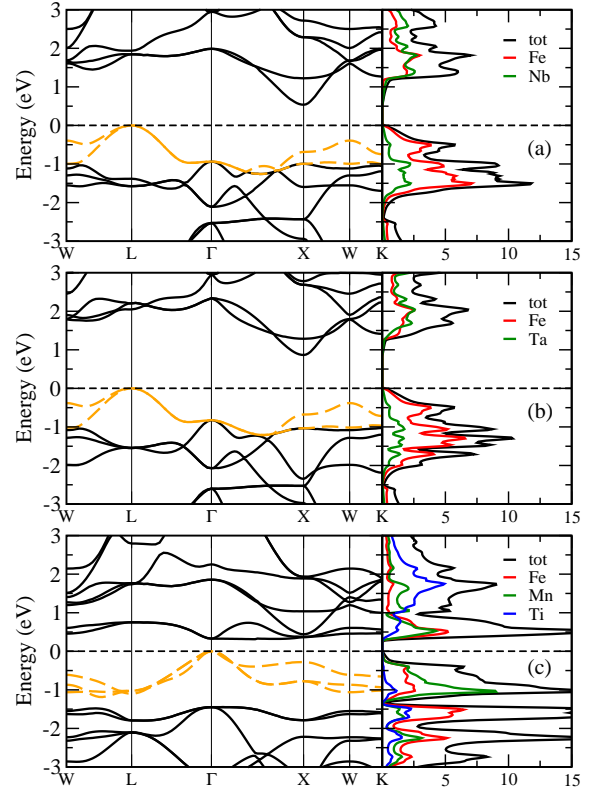}
 \caption{Calculated electronic structures of (a) FeNbSb, (b) FeTaSb, and (c) FeMnTiSb, in cubic F$\bar{4}$3m symmetry. The top of the valence band is taken as zero on the energy axis.}
 \label{electronic}
\end{figure}

The electronic structures of FeNbSb, FeTaSb, and FeMnTiSb were calculated using GGA-PBEsol. The band structure and density of states (DOS) are shown in Fig.~\ref{electronic}. The electronic features of FeNbSb and FeTaSb are quite similar owing to the similar size and chemical properties of Nb and Ta. This is also reflected in the electrical transport properties discussed later. The valence band maximum (VBM) is at \textit{L}-point whereas the conduction band minimum (CBM) is at \textit{X}-point for both FeNbSb and FeTaSb i.e., indirect band gap semiconductor. 

The VBM of FeNbSb and FeTaSb is twofold degenerate and comprises of heavy and light bands. The heavy band enhances Seebeck coefficient whereas the light band facilitates the mobility of charge carriers. Thus, the combination of heavy and light bands is helpful in achieving good TE performance \cite{Gudelli13, Zhang10, Yang14}. The DOS features show that the VBM of FeTaSb mostly comprises of \textit{d}-states of Fe and some \textit{d}-states of Ta, very similar to FeNbSb \cite{Li16, Zhang16}. The CBM is dominated by \textit{d}-states of Fe and Ta. There is no contribution whatsoever from the Sb states. Hence, the doping at Sb site may improve the carrier concentration without affecting the band structure. Any doping at Ta position will be reflected in the CBM without affecting the VBM significantly. However, doping at Fe site will be reflected in both VBM and CBM. These theoretical considerations could be helpful for choosing the experimental doping levels. The energy dispersion of the bands suggests a \textit{p}-type behavior, which is indeed confirmed by electrical transport properties \cite{Zhang16}. 
 
In case of FeMnTiSb, the VBM is threefold degenerate whereas the CBM is twofold degenerate. The nature of band gap is direct and VBM and CBM are located at \textit{$\Gamma$}-point. The VBM comprises of heavy bands which are relatively flat in \textit{$\Gamma$-X} direction than those of FeNbSb and FeTaSb. Flat bands are an indication of large effective mass which augments the Seebeck coefficient. But, a large effective mass reduces the mobility of charge carriers, thereby decreasing the electrical conductivity. These proposed effects have been observed in the electrical transport calculations discussed ahead: the Seebeck coefficient of FeMnTiSb is larger than that of FeTaSb whereas the electrical conductivity of FeMnTiSb is lower than that of FeTaSb, in both \textit{n}-type and \textit{p}-type doping. 

The CBM of FeMnTiSb is twofold degenerate and comprises of flat bands. One of the bands is almost flat in \textit{$\Gamma$-X}, \textit{X-W}, and \textit{W-K} directions. These less dispersed bands at CBM of FeMnTiSb suggest improvement in TE performance on \textit{n}-type doping as is confirmed in the coming section \cite{Yadav15}. The DOS features show that the VBM of FeMnTiSb comprises mostly of Mn \textit{d}-states, with some contribution from the \textit{d}-states of Fe and Ti. The CBM is dominated by the similar contribution from the \textit{d}-states of Fe and Mn. Yet again, no significant contribution of Sb states at either VBM or CBM. Therefore, the doping at Sb sites may be helpful in improving the TE properties without altering the band structure. Doping at Mn site will affect considerably both VBM and CBM, whereas doping at Fe position will significantly influence only the CBM of FeMnTiSb. Next, we proceed to see the effect of discussed electronic features on electrical transport properties.
  
\subsection{Thermoelectric Parameters}
The thermoelectric properties of all the systems are calculated using rigid band approximation (RBA), semiclassical Boltzmann theory, and constant relaxation time approach. The RBA assumes that on doping a system, Fermi level moves up or down without affecting the band structure. Thus, a single band structure calculation is sufficient for all the doping levels. The RBA holds good for low doping levels and has been widely employed \cite{Lee11, Madsen06, Chaput05, Jodin04}. In this section, first, we predict the trend of PF with doping for FeNbSb and further compare our results with experimentally reported ones. Then, we discuss the TE properties of FeTaSb and FeMnTiSb. We have chosen three representative temperatures namely 300 K, 700 K, and 1100 K for the evaluation of TE properties. The purpose is to see whether the material is compatible with room temperature or conventional high temperature TE applications or for both. For convenience, throughout this paper, we use \textit{p}-FeNbSb, \textit{p}-FeTaSb, \textit{p}-FeMnTiSb for \textit{p}-type doped systems, and \textit{n}-FeMnTiSb for \textit{n}-type doped system.
  
\begin{figure}
\centering
 \includegraphics[scale=0.4]{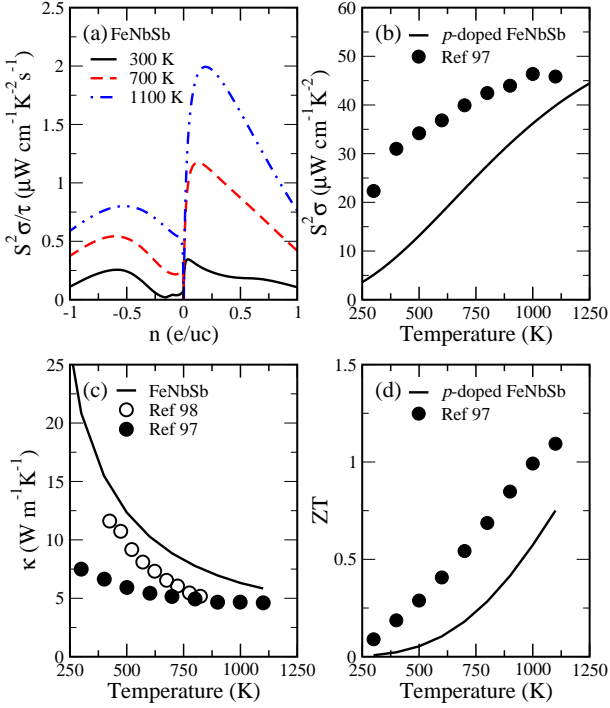}
 \caption{(a) Calculated $S^2\sigma$/$\tau$ as a function of doping (per unit cell) for FeNbSb at 300 K, 700 K, and 1100 K. (b) Calculated and reported $S^2\sigma$ as a function of temperature for 0.2 \textit{p}-type doped FeNbSb (c) Thermal conductivity versus temperature for calculated FeNbSb, reported FeNbSb (empty circles), and reported 0.2 \textit{p}-type doped FeNbSb (filled circles) (d) The figure of merit as a function of temperature for calculated and reported 0.2 \textit{p}-type doped FeNbSb.}
\label{pf}
\end{figure}

The calculated PF/\textit{$\tau$} as a function of doping (per unit cell) at different temperatures for FeNbSb is shown in Fig.~\ref{pf}(a). The trend of PF/\textit{$\tau$} is increasing for both \textit{n}-type and \textit{p}-type doping and then falls gradually at higher doping levels. The behavior is consistent at all considered temperatures. But the maximum PF/\textit{$\tau$} is achieved for \textit{p}-type doping at all the temperatures. This is in good agreement with previously calculated and experimental results \cite{Fu15, Fu16, Yu17, Li16, Zhang16}. The optimal \textit{p}-type doping levels for maximum PF/\textit{$\tau$} are 0.04, 0.12, and 0.20, at 300 K, 700 K, and 1100 K, respectively. Yang \textit{\textit{\textit{\textit{\textit{et al.}}}}} also proposed 0.04 as the optimal \textit{p}-type doping level for FeNbSb at room temperature \cite{Yang08}. Further, the maximum PF/\textit{$\tau$} at 1100 K for 0.2 \textit{p}-type doping is in good agreement with experimental results. Fu \textit{\textit{\textit{\textit{\textit{et al.}}}}} reported a peak \textit{ZT} of 1.1 at 1100 K for FeNb$_{0.8}$Ti$_{0.2}$Sb i.e., 0.2 \textit{p}-type doping \cite{Fu15RSC}. 

A comparison of calculated and experimental PF of \textit{p}-FeNbSb is shown in Fig.~\ref{pf}(b). To compare the theory with experiment, we use an approximate value of relaxation time i.e., \textit{$\tau$} = 2 x 10$^{-15}$ s, arrived at by comparing the calculated and experimental electrical conductivity of \textit{p}-FeNbSb. Using \textit{$\tau$} = $\sigma_{exp}$/$\sigma_{cal}$ suggests that \textit{$\tau$} is of the order of 10$^{-15}$ s. The strategy has been helpful in our previous work and we obtained a reasonable agreement for cobalt based 18 VEC systems \cite{Zeeshan17}. Thus, the adopted value of relaxation time is a realistic representative value. The trend of calculated PF is quite similar to that of the experimental PF for \textit{p}-FeNbSb. The difference in magnitude is expected since our calculations are for the pristine system under ideal conditions, very unlikely to be the case for experimental study. The magnitude may vary more or less depending on the system under consideration and techniques employed. However, the trend of the two curves is more important for us to predict the properties of materials. 
 
Fig.~\ref{pf}(c) shows calculated thermal conductivity as a function of temperature for base composition FeNbSb, along with reported data on FeNbSb and \textit{p}-FeNbSb. Thermal conductivity is comprised of two components, electronic \textit{$\kappa_e$}, and lattice \textit{$\kappa_l$}. The additivity of the electron and lattice heat currents for the thermal conductivity calculations result from linear response theory and may physically be attributed to the fact that the electron distribution under thermal current conditions has in leading order the same scattering rate for the phonons as when under equilibrium conditions.

The dominance of \textit{$\kappa_l$} in total thermal conductivity is well known for Heusler alloys \cite{He14, Wu09, Kimura08}. The plots for calculated and reported \textit{$\kappa$} are in good agreement for the parent material FeNbSb in the range of 400-825 K\cite{Tavassoli17}. The calculated values slightly overestimate the reported ones. Hence we need to keep in mind that our calculations would yield an underestimated value of the figure of merit. The experimental data shows that the \textit{$\kappa$} of reported \textit{p}-FeNbSb has lower values than that of parent FeNbSb as expected on account of mass fluctuations. Similar to PF, the calculated \textit{$\kappa$} for FeNbSb approaches the \textit{$\kappa$} of reported \textit{p}-FeNbSb at the higher temperature. Note that the calculated values of \textit{$\kappa$} are lowered by tenfold. This is our observation from the previous study that the calculated values of \textit{$\kappa_l$} by QUANTUM ESPRESSO are, in general, tenfold higher\cite{Zeeshan2017}.
  
\begin{figure}
\centering
 \includegraphics[scale=0.4]{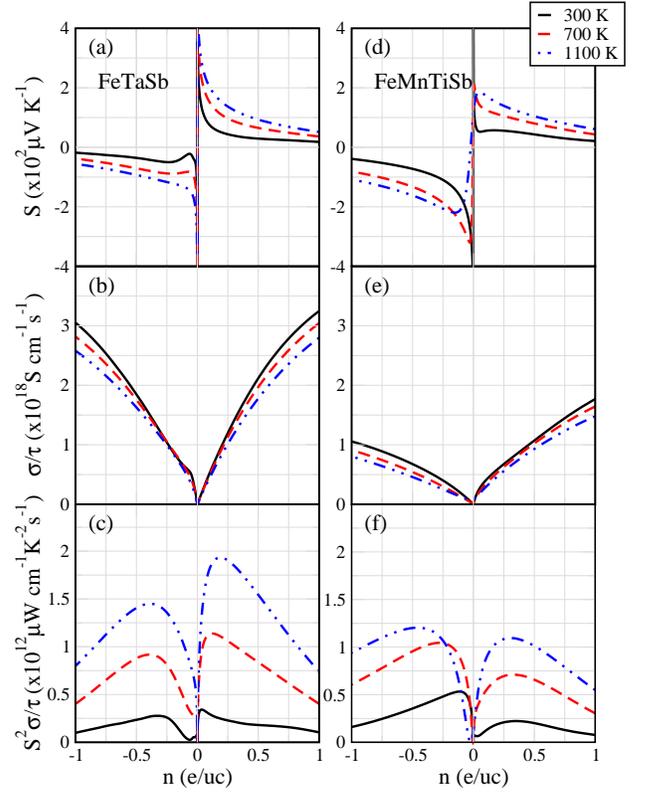}
 \caption{The trend of Seebeck coefficient, electrical conductivity, and power factor as a function of doping per unit cell at 300 K, 700K, and 1100 K, for FeTaSb (a-c) and FeMnTiSb (d-f). Power factor and electrical conductivity are plotted with respect to relaxation time.}
\label{transport}
\end{figure}

The behavior of figure of merit \textit{ZT} as a function of temperature for calculated and reported \textit{p}-FeNbSb, illustrated in Fig.~\ref{pf}(d) \cite{Fu15RSC}, exhibits similar trend over the entire temperature range. However, the calculated values lag behind the experimental ones. It is noteworthy here that the difference in the magnitude of the two can be attributed to the fact that we employed the \textit{$\kappa$} of undoped FeNbSb for calculating the \textit{ZT} of \textit{p}-FeNbSb. As can be seen from Fig.~\ref{pf}(c), (i) our calculations overestimate \textit{$\kappa$} w.r.t. the experiment, (ii) a significant reduction in \textit{$\kappa$} is observed with doping. Consequently, we expect our calculated \textit{ZT} values to be much lower than the expected ones. The calculations of \textit{$\kappa$} for doped systems are relatively complicated and computationally demanding, therefore, we resorted to adopting \textit{$\kappa$} of undoped FeNbSb for the doped material also. Nevertheless, we obtain similar trends for calculated and reported TE properties of \textit{p}-FeNbSb. As stated earlier, the trend is more important for us rather than the actual figures. Moreover, the close proximity of the calculated and reported plots for PF and \textit{$\kappa$} at higher temperatures imply that our calculated results are comparable to reported ones in that range. Having judged our results for the reference material FeNbSb, in the following we present our findings on the two proposed TE systems FeTaSb and FeMnTiSb.

The dependence of TE parameters on doping at different temperatures for FeTaSb and FeMnTiSb is illustrated in Fig.~\ref{transport}. The Seebeck coefficient, Fig.~\ref{transport}(a) and ~\ref{transport}(d), is the maximum, for both the materials, when the Fermi level is near the middle of the band gap and decreases at higher doping levels, showing almost the same trend for all the temperatures under consideration. Exactly an opposite trend is observed for electrical conductivity which rises  rapidly with doping, Fig.~\ref{transport}(b) and ~\ref{transport}(e). As discussed under electronic features, the Seebeck coefficient of FeMnTiSb is higher than that of FeTaSb whereas the electrical conductivity of FeMnTiSb is lower than that of FeTaSb owing to heavy flat bands at VBM. As the Fermi level shifts towards the VBM or CBM, the carrier concentration increases, thereby improving the electrical conductivity and reducing the Seebeck coefficient. Therefore, an optimal doping level is desired which maintains a balance between Seebeck coefficient and electrical conductivity to attain a maximum PF. Generally, such a doping level is found when the Fermi level is near the band edge \cite{Yadav15, Lee11, Parker10PRB, Yang08}. 
 
\begin{figure}
\centering
 \includegraphics[scale=0.4]{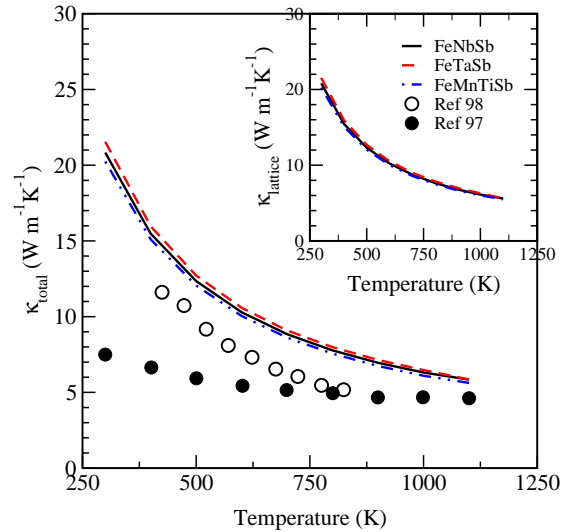}
 \caption{The calculated total thermal conductivity of FeNbSb, FeTaSb, FeMnTiSb, and reported total thermal conductivity of FeNbSb and 0.2 \textit{p}-type doped FeNbSb. The inset shows the lattice thermal conductivity of FeNbSb, FeTaSb, and FeMnTiSb.}
\label{thermal}
\end{figure}

The calculated PF/\textit{$\tau$} as a function of doping (per unit cell) at different temperatures for FeTaSb and FeMnTiSb is shown in Fig.~\ref{transport}(c) and ~\ref{transport}(f). The magnitude of PF/\textit{$\tau$} enhances with doping, and exhibits quite a similar behavior for both the systems: first increasing sharply with doping and then falling steadily at higher doping levels. The \textit{p}-type doping dominates \textit{n}-type at all the temperatures for FeTaSb. The optimal \textit{p}-type doping levels for best PF/\textit{$\tau$} values are 0.03, 0.11, and 0.18, at 300 K, 700 K, and 1100 K, respectively. However, the \textit{n}-type doping prevails in case of FeMnTiSb at different temperatures. The maximum PF/\textit{$\tau$} values are obtained at 0.11, 0.27, and 0.47 \textit{n}-type doping at 300 K, 700 K, and 1100 K, respectively. The doping levels are quite pragmatic and could be realized experimentally.

The hole doped FeNbSb has recently been reported to have the highest \textit{ZT} among the hH alloys \cite{Yu17}. It is encouraging to note that the PF/\textit{$\tau$} for the proposed material \textit{p}-FeTaSb at 1100 K is comparable to that of \textit{p}-FeNbSb. This shows the potential of FeTaSb as a high temperature TE material.

Now we focus our attention on the quaternary alloy FeMnTiSb being explored by us for TE properties. It is highly desirable to have both \textit{n}-type and \textit{p}-type branches of a TE module to be of similar TE materials or ideally of the same material. Efficient combination of \textit{n}-type and \textit{p}-type branches of the same material is rare. In most of the cases, the performance of one type is poor in comparison to the other\cite{Dubois_Book}. Interestingly, at 1100 K, the values of maximum PF/\textit{$\tau$} for \textit{n}-type and \textit{p}-type doped FeMnTiSb are very close; the corresponding optimal doping levels being 0.47 and 0.31, respectively. This prompts us to propose the importance of FeMnTiSb as both \textit{n}-type and \textit{p}-type high temperature TE material.

Another noteworthy observation from Fig.~\ref{transport}(f) is that the maximum calculated PF/\textit{$\tau$} at 300 K of FeMnTiSb is about 1.5 times that of recently much popular or acclaimed FeNbSb. Thus, FeMnTiSb could be a better room temperature TE prospect in comparison to FeNbSb and FeTaSb. High PF/\textit{$\tau$} values for both \textit{n}-type and \textit{p}-type FeTaSb at 1100 K (Fig.~\ref{transport}(c)) reveal that FeTaSb may also fall in the category of materials which can be utilized for making both \textit{n}-type and \textit{p}-type branches. The discussion so far indicates that the PF of \textit{p}-FeTaSb is comparable with \textit{p}-FeNbSb whereas \textit{n}-FeMnTiSb could be a better room temperature TE prospect. Therefore, it becomes interesting to see the behavior of their thermal conductivity.
 
\begin{table}[]
\centering
\setlength{\arrayrulewidth}{0.5pt}
\begin{tabular}{l|c|c|c|c|c|c}
\hline
\hline
System		     & T     & n      &c    & S     & PF   & ZT    \\ \hline
FeNbSb              & 300   & 0.04   &0.7  & 127   & 5.1  & 0.007 \\
                     & 700   & 0.12   &2.4  & 139   & 22.7 & 0.179  \\
                     & 1100  & 0.20   &3.7  & 155   & 39.8 & 0.749  \\
\hline
FeNbSb (exp.)\cite{Fu15RSC} & 1100  & 0.20   & & 203   & 45.8 & 1.1   \\	
\hline
FeTaSb               & 300   & 0.03   &0.7  & 131   & 6.8  & 0.009 \\
                     & 700   & 0.11   &2.3  & 140   & 22.6 & 0.174 \\
                     & 1100  & 0.18   &3.6  & 156   & 38.6 & 0.725 \\
\hline
FeMnTiSb             & 300   & -0.11  &2.0  & 162   & 10.6 & 0.030  \\
                     & 700   & -0.27  &5.0  & 180   & 20.4 & 0.169 \\ 
                     & 1100  & -0.47  &8.8  & 166   & 24.0 & 0.469 \\ \hline \hline
\end{tabular}
\caption{The calculated optimal doping levels (n, per unit cell) for \textit{p}-type FeNbSb, \textit{p}-type FeTaSb, and \textit{n}-type FeMnTiSb are shown at different temperatures (T, in K) and the corresponding carrier concentration (c, in 10$^{21}$ cm$^{-3}$), Seebeck coefficient(S, in $\mu$V K$^{-1}$), power factor (PF, in $\mu$W cm$^{-1}$ K$^{-2}$), and \textit{ZT}. The reported values of FeNbSb at 1100 K are also listed for comparison.}
\label{doping}
\end{table}

Fig.~\ref{thermal} shows the trend of calculated \textit{$\kappa$} with temperature for FeNbSb, FeTaSb, and FeMnTiSb. For comparison, reported \textit{$\kappa$} of parent FeNbSb and \textit{p}-FeNbSb are also included. The inset establishes the dominance of \textit{$\kappa_l$} in total \textit{$\kappa$}. As discussed earlier, the trend of calculated and reported \textit{$\kappa$} of parent FeNbSb is quite similar. However, the calculated values slightly overestimate the reported ones. The effect of doping is quite pronounced on the \textit{$\kappa$} of \textit{p}-FeNbSb. Importantly, the two curves for reported \textit{p}-doped and calculated undoped FeNbSb approach each other at the higher temperature. Despite the low projected values, the \textit{$\kappa$} of FeTaSb and FeMnTiSb bear a close resemblance to that of FeNbSb over the entire temperature range. We, as stated earlier, expected the \textit{$\kappa$} of quaternary FeMnTiSb to be lower than that of ternary hH alloys, however, the reduction in \textit{$\kappa$} is not substantial.

For a quantitative analysis, Table~\ref{doping} presents the optimal doping levels and the corresponding Seebeck coefficient, PF, and \textit{ZT} for FeTaSb and FeMnTiSb, \textit{vis-a-vis} those of the reference material FeNbSb from experiments and our calculations. The values of Seebeck coefficient range from 127-186 $\mu$V K$^{-1}$ and are in line with many high performance hH alloys. The optimal doping concentrations are of the order of 10$^{21}$ cm$^{-3}$ which are most suitable for obtaining a high TE performance by hH alloy. 

As discussed earlier, assuming a relaxation time \textit{$\tau$} = 2 x 10$^{-15}$ s, we propose an approximate value of PF and \textit{ZT} for FeTaSb and FeMnTiSb at optimal doping concentrations. The calculated and reported PF for \textit{p}-FeNbSb \cite{Fu15RSC},  at 1100 K, are 39.8 $\mu$W cm$^{-1}$ K$^{-2}$ and 45.8 $\mu$W cm$^{-1}$ K$^{-2}$, respectively. Thus, the good agreement between experimental values and our results indicates the reliability of our calculations and the suitability of the chosen relaxation time, especially at high temperatures. At 1100 K, the calculated PF of \textit{p}-FeTaSb is 38 $\mu$W cm$^{-1}$ K$^{-2}$ which is again comparable to calculated and reported PF of \textit{p}-FeNbSb. Whereas FeTaSb could be as challenging as FeNbSb at high temperatures, the PF of \textit{n}-FeMnTiSb at 1100 K is still comparable with conventional TE material CoTiSb, reported \cite{Wu07, Qiu09} to have the maximum PF of about 23.2 $\mu$W cm$^{-1}$ K$^{-2}$. The most remarkable observation from Table~\ref{doping} is that the \textit{n}-FeMnTiSb shows a high PF of 10.6 $\mu$W cm$^{-1}$ K$^{-2}$ at room temperature which is higher than that of FeNbSb. Despite  being a low value, this PF is still competitive enough with that of many CoTiSb based materials \cite{Wu07, Qiu09}.

Incorporating \textit{$\kappa$} of parent systems, displayed in Fig.~\ref{thermal}, in the \textit{ZT} formula for \textit{p}-FeNbSb, \textit{p}-FeTaSb, and \textit{n}-FeMnTiSb, the obtained underestimated, as discussed earlier, \textit{ZT} values are listed in Table~\ref{doping}. The calculated and reported \textit{ZT} values at 1100 K for \textit{p}-FeNbSb are 0.74 and 1.1, respectively. Remarkably, the figure of merit \textit{ZT} $\sim$ 0.72 for \textit{p}-FeTaSb at 1100 K, is not far behind from that of \textit{p}-FeNbSb, thus ensuring that \textit{p}-FeTaSb is as promising as \textit{p}-doped FeNbSb. It is worth stressing here that, as discussed earlier, the actual values of the figure of merit could be as high as unity for both \textit{p}-FeNbSb and \textit{p}-FeTaSb provided \textit{$\kappa$} of doped FeNbSb and FeTaSb is included in the \textit{ZT} formula. 
 
The calculated \textit{ZT} of \textit{n}-FeMnTiSb at room temperature is quite low. The reason is quite obvious. The \textit{$\kappa$} of undoped FeMnTiSb is fairly high at room temperature (Fig.~\ref{thermal}). Since the \textit{$\kappa$} of FeMnTiSb parallels \textit{$\kappa$} of FeNbSb, we expect that the doped FeMnTiSb would have about threefold lower \textit{$\kappa$} at room temperature, thereby improving the overall figure of merit. Nevertheless, the calculated \textit{ZT} values of \textit{p}-FeMnTiSb and \textit{n}-FeMnTiSb, at 1100 K, are 0.42 (not shown in Table) and 0.46, respectively. The values are lower than that of FeNbSb but compatible enough with CoTiSb based materials\cite{Wu07, Qiu09}. Similar to FeNbSb and FeTaSb, the actual \textit{ZT} values of \textit{n}-FeMnTiSb and \textit{p}-FeMnTiSb are expected to be higher on the inclusion of \textit{$\kappa$} of doped FeMnTiSb in the figure of merit formula.
 
\section{Conclusions}

In this work, utilizing \textit{ab initio} approach, semiclassical Boltzmann transport theory, and constant relaxation time approach, we have systematically investigated the ground state properties, structural stability, electronic features (band structure and DOS), and thermal and electrical transport properties of two novel Fe-based Heusler alloys FeTaSb and FeMnTiSb. Both the systems are nonmagnetic semiconductors in cubic {F$\bar{4}$3m} symmetry and their stability is confirmed by phonon calculations. At 1100 K, the power factor of \textit{p}-doped FeTaSb (38.6 $\mu$W cm$^{-1}$ K$^{-2}$) is comparable to the best performing Heusler alloy FeNbSb whereas the power factor of \textit{n}-doped FeMnTiSb (24 $\mu$W cm$^{-1}$ K$^{-2}$) is comparable to the conventional TE material CoTiSb. However, the power factor of FeMnTiSb (10.6 $\mu$W cm$^{-1}$ K$^{-2}$) at room temperature is higher than both FeNbSb and FeTaSb. Interestingly, at high temperatures, the low cost FeMnTiSb could be used as both \textit{n}-type and \textit{p}-type legs in a thermoelectric module. The \textit{ZT} at 1100 K of \textit{p}-doped FeTaSb is 0.72 whereas \textit{ZT} of \textit{n}-doped and \textit{p}-doped FeMnTiSb are 0.46 and 0.42, respectively, which, as discussed, are actually underestimated values. We are optimistic that our findings, suggesting the potential of FeTaSb and FeMnTiSb as promising thermoelectric materials, would motivate and prompt the experimentalists to realize these materials and their potential. 

\section{Acknowledgements}
M.Z. is thankful to CSIR for the support of a senior research fellowship. Computations were performed on HP cluster at the Institute Computer Center (ICC), IIT Roorkee, and at IFW Dresden, Germany. We thank Navneet Gupta and Ulrike Nitzsche for technical assistance. H.C.K. gratefully acknowledges financial support from the FIG program of IIT Roorkee (Grant CMD/FIG/100596).

\end{document}